**Diffusion Dispersion Imaging: Mapping OGSE Frequency Dependence in the Human Brain**

Aidin Arbabi[1], Jason Kai[1], Ali R. Khan[1], and Corey A. Baron[1]

1. Centre for Functional and Metabolic Mapping, Robarts Research Institute, Schulich School of Medicine & Dentistry, The University of Western Ontario, London, Ontario, Canada

*Corresponding Author*

Corey A Baron

Email: cbaron@robarts.ca

Address: 1151 Richmond St. N., Dock 50, London, ON, N6A 5B7, Canada

Telephone: 1-519-931-5777 Ext. 24420

# Submitted to Magnetic Resonance in Medicine

**Abstract**

*Purpose*: Oscillating gradient spin-echo (OGSE) diffusion MRI provides information about the microstructure of biological tissues via the frequency dependence of the apparent diffusion coefficient (ADC). ADC dependence on OGSE frequency has been explored in numerous rodent studies, but applications in the human brain have been limited and have suffered from low contrast between different frequencies, long scan times and a limited exploration of the nature of the ADC dependence on frequency.

*Methods*: Multiple frequency OGSE acquisitions were acquired in healthy subjects at 7 T to explore the power-law frequency dependence of ADC, the "diffusion dispersion". Further, a method for optimizing the estimation of the ADC difference between different OGSE frequencies was developed, which enabled the design of a highly efficient protocol for mapping diffusion dispersion.

*Results*: For the first time, evidence of a linear dependence of ADC on the square root of frequency in healthy human white matter was obtained. Using the optimized protocol, high quality, full-brain maps of apparent diffusion dispersion rate were also demonstrated at an isotropic resolution of 2 mm in a scan time of 6 minutes.

*Conclusion*: This work sheds light on the nature of diffusion dispersion in the healthy human brain and introduces full brain diffusion dispersion mapping at clinically relevant scan times. These advances may lead to new biomarkers of pathology or improved microstructural modelling.

**Introduction**

Water diffusion in biological tissues is restricted by microstructure composition. As a result, the apparent diffusion coefficient (ADC) measured with diffusion magnetic resonance imaging (dMRI) generally depends on the effective diffusion time ($\Delta_{eff}$), the time during which water molecules probe their surrounding environment. As diffusion times approach zero, molecules only travel short distances and fewer interact with barriers like cellular membranes, and the estimated ADC approaches the intrinsic diffusion coefficient up to surface-to-volume effects (1,2). For longer diffusion times, on the other hand, water spins have a higher chance of interacting with obstacles and the observed ADC will be decreased. Thus, measuring $\Delta_{eff}$-dependence of ADC provides an opportunity for additional insight into the microstructure of biological tissues compared to ADC alone.

Traditional dMRI is performed using pulsed gradient spin-echo (PGSE) with $\Delta_{eff}$ typically greater than 30 ms in human applications (3). In Stepisnik's groundbreaking work in the 1980's, oscillating gradient spin-echo (OGSE) encoding was introduced as a method that enables short $\Delta_{eff}$ by using rapidly oscillating diffusion gradients, as $\Delta_{eff}$ scales inversely with oscillating frequency, ω (4,5). OGSE diffusion encoding provides an extra dimension for probing axon diameter (6), surface-to-volume ratios (7) and microstructural disorder (8), and has been demonstrated to provide unique sensitivity to microstructural changes in pathology for several preclinical studies. For example, Does *et al.* studied the ω-dependence of ADC in the gray matter of normal and globally ischemic rat brain with frequencies ranging from 0 ($\Delta_{eff}$ = 10 ms) to 1000 Hz, and observed ADC increases as much as 24% in vivo and 50% postmortem (9). Bongers et al. found that OGSE was more effective than PGSE as an early MRI biomarker for radiation therapy response monitoring in glioblastoma mouse models, and that tumor ADC was generally 30-50% higher than in surrounding white matter for a frequency of 200 Hz compared to 0 Hz ($\Delta_{eff}$ = 18 ms) (10). They also detected a 15% increase in the tumor ADC in response to radiation, while PGSE showed a lower sensitivity to radiation changes. Gore et al. also showed OGSE is a potentially earlier and more sensitive indicator of tumor treatment response than conventional PGSE (11). Potential benefits of using OGSE encoding in delineating tissue microstructure has also been reported in other studies of animal models of stroke (12), multiple sclerosis (13), and cancer (14). High performance small-bore systems have also recently enabled the combination of OGSE and multiple diffusion encoding, which may provide a new dimension of sensitivity to investigate pathology (15,16).

The successful application of OGSE encoding and the unique insight into pathology it enables in animal models makes its translation into human studies appealing. However, lower gradient strengths on human MR systems significantly reduces the maximum attainable $b$-value and frequency for a given echo time. Consequently, *in vivo* human OGSE acquisitions suffer from an inherently low ADC-to-noise ratio (17). Nevertheless, ADC dependence on OGSE frequency has been observed in both grey and white matter regions in the healthy human brain (18,19), and OGSE can provide complementary microstructural information to PGSE in acute ischemic stroke (20). However, scan times were long (20 minutes for full brain coverage with 2.5 mm thick slices), single-voxel maps of ADC differences between PGSE and OGSE have had poor SNR, and a parameterization of the dependence of ADC on frequency (i.e., the "diffusion dispersion" (8)) has not been demonstrated in the in vivo human brain.

Notably, a $\omega^\theta$ dependence of ADC has been predicted both in the short ($\omega \rightarrow \infty$) and long ($\omega \rightarrow 0$) diffusion time regimes (21). In the short diffusion time regime, $\theta$ = -1/2 and ADC differences are directly proportional to surface-to-volume ratios (2,7). In the long diffusion time regime, coarse graining occurs and the dependence on frequency is related to long-range structural correlations, where $\theta$ is a parameter given by the effective dimension of diffusion and the class of structural disorder (8). $\theta$ = 1/2 has been demonstrated in both healthy (22) and globally ischemic (8,9) rodent brain tissue. A trend towards $\theta < 1$ can be observed from the data presented in the in vivo human brain (18), but this behaviour was not explicitly explored and only 2 non-zero frequencies were acquired.

In this work, we explored the $\omega^\theta$ dependence of ADC in healthy subjects for frequencies in the range of 0 to 60 Hz by performing in vivo PGSE and OGSE ADC mapping at 7 T with $b$ = 450 s/mm². For the first time, evidence for θ = 1/2 has been obtained in the in vivo human brain. Capitalizing on this finding, an optimized protocol was developed to acquire high SNR, clinical-resolution (2 mm isotropic) full-brain maps of the ADC difference between PGSE and OGSE in a scan time of only 6 minutes.

**Theory**

*Mapping ADC Differences:* Considering a power law relationship between ADC and OGSE frequency (8), we define the apparent diffusion dispersion rate ($\Lambda$) as the slope of linear regression of ADC with $\omega^\theta$:

$$D_\omega = \Lambda \omega^\theta + D_{\omega 0} \qquad [1]$$

where $D_\omega$ is the OGSE ADC at a frequency $\omega$ and $D_{\omega 0}$ is the ADC at $\omega = 0$. Accordingly, the apparent diffusion dispersion rate is directly proportional to the difference in ADC between an OGSE ($\omega > 0$) and PGSE ($\omega \approx 0$) scan, $\Delta D$:

$$\Lambda = \frac{\Delta D}{\omega^\theta} \qquad [2]$$

Thus, mapping $\Delta D$ can serve as a surrogate for mapping the apparent diffusion dispersion rate that requires only a single OGSE and PGSE acquisition. Accordingly, considering computation of the mean ADC (MD) from a uniformly distributed multi-directional acquisition (e.g., tetrahedral encoding (17,23)) for PGSE and OGSE at a single frequency, the expression for $\Delta D$ is:

$$\Delta D = -\frac{1}{N_\omega} \sum_{i=1}^{N_\omega} \frac{ln\left(\frac{S_{f,i}}{S_0}\right)}{b_{\omega,i}} + \frac{1}{N_{\omega 0}} \sum_{i=1}^{N_{\omega 0}} \frac{ln\left(\frac{S_{\omega 0,i}}{S_0}\right)}{b_{\omega 0,i}} \qquad [3]$$

where $N_\omega$ and $N_{\omega 0}$ are the number of OGSE and PGSE acquisitions, respectively, $S_0$ is the $b$ = 0 signal, $S_{\omega,i}$ and $S_{\omega 0,i}$ are the direction-dependent diffusion weighted signals at frequencies $\omega > 0$ and $\omega = 0$, respectively, and $b_{\omega,i}$ and $b_{\omega 0,i}$ are the direction-dependent $b$-values at frequencies $\omega > 0$ and $\omega = 0$, respectively.

While the $b$-values would ideally be identical for all acquisitions, small differences in $b$ will likely occur in practice due to cross terms that arise from the crusher gradients on either side of the refocusing RF pulse. However, assuming these variations in $b$ are small, the expression can be simplified to a format where a $b = 0$ acquisition is not required, which is advantageous for scan time reductions that could help facilitate clinical translation:

$$\Delta D = \frac{1}{N_\omega} \sum_{i=1}^{N_\omega} \frac{-ln(S_{\omega,i})}{b_{\omega,i}} + \frac{1}{N_{\omega 0}} \sum_{i=1}^{N_{\omega 0}} \frac{ln(S_{\omega 0,i})}{b_{\omega 0,i}} + E \qquad [4]$$

$$E = \frac{1}{N_\omega} \sum_{i=1}^{N_\omega} \frac{ln(S_0)}{b_{\omega,i}} - \frac{1}{N_{\omega 0}} \sum_{i=1}^{N_{\omega 0}} \frac{ln(S_0)}{b_{\omega 0,i}} \qquad [5]$$

where $E$ is a bias incurred by omitting the acquisition of $b = 0$ images. $E = 0$ when identical $b$-values are used for all acquisitions. Notably, for *a priori* known $\theta$, the apparent diffusion dispersion rate can be readily determined from $\Delta D$ using Eq. 2.

*ADC Difference Map Optimization*: To optimize a protocol for ADC differences, sequence parameters that maximize the ratio of the mean $\Delta D$ to its standard deviation can be evaluated, similar to approaches that have been used to determine optimal parameters for the measurement of ADC (24). For these purposes, identical $b$-values for all directions and frequencies (i.e., $E = 0$) and a direction-independent diffusion tensor are assumed in Eq. 4, which leads to the expression (Appendix A):

$$\frac{\Delta D}{\sigma_{\Delta D}} = SNR_0 \sqrt{N_{\omega 0}} \cdot b \Lambda \omega^\theta \left(1 + \frac{e^{2b\Lambda \omega^\theta}}{R_{op}}\right)^{-\frac{1}{2}} e^{-bD_{\omega 0}} e^{-TE(b)/T_2} \qquad [6]$$

where $\sigma_{\Delta D}$ is the standard deviation of estimated ADC difference between OGSE and PGSE, $SNR_0$ is the signal-to-noise ratio of a proton density scan with TE = 0 and b = 0, TE(b) is the $b$

-value dependent echo-time, $D_{\omega 0}$ is the PGSE ADC value, $N_{\omega 0}$ is the number of PGSE acquisitions, and $R_{op}$ is the ratio of the number of OGSE to PGSE acquisitions. It can be shown that $\Delta D/\sigma_{\Delta D}$ is maximized when $R_{op} = e^{b\Lambda\omega^\theta}$ (Appendix A), which together with Eq. 6 can be used to determine the diffusion encoding parameters ($b$, $f$, $TE$, and $R_{op}$) that maximize $\Delta D/\sigma_{\Delta D}$ for typical $\theta$, $\Lambda$, $D_{\omega 0}$, and $T_2$.

**Methods**

*Multiple Frequency OGSE*: MRI scans were performed in a water phantom and 6 healthy male subjects on a 7 T head-only system (80 mT/m strength and 350 T/m/s slew rate). This study was approved by the Institutional Review Board at Western University, and informed consent was obtained prior to scanning. To mitigate eddy current artefacts and reduce acoustic noise and gradient duty cycle, the maximum gradient was limited to 68 mT/m with 240 T/m/s slew rate for OGSE scans. Multiple frequency dMRI data was acquired in a single scan that employed standard PGSE ($\Delta_{eff}$ = 41 ms, 0 Hz) and cosine-modulated trapezoidal OGSE with frequencies 30 Hz, 45 Hz, and 60 Hz (Figure 1). The remaining parameters were $b$ = 450 s/mm², 4 direction tetrahedral encoding and $b$ = 0 acquisitions with 10 averages each, $TE/TR$ = 111/5500 ms, $FOV$ = 200 × 200 mm², 2.5 mm isotropic in-plane resolution, 32 slices (3 mm), and scan time 18 min. The image volume was interpolated to 1.25 mm × 1.25 mm × 1.50 mm resolution before analysis. Signal changes with respect to OGSE frequency are relatively small (17,18) and estimation of $\Lambda$ and $\theta$ may be particularly sensitive to imaging artefacts compared to ADC. Accordingly, in this initial work parallel imaging was not implemented to mitigate residual aliasing artefacts and to maximize SNR. Registration between diffusion directions and frequencies was performed using FSL (25).

For anatomical reference, $b$ = 1000 s/mm² standard PGSE diffusion tensor imaging (DTI) was acquired with 30 directions and 6 $b$ = 0 acquisitions (TE/TR = 53/8200 ms, FOV = 200 × 200 mm², 2 mm isotropic resolution, scan time 5 min). The DTI scan was registered to the OGSE scan, which was followed by probabilistic whole-brain tractography using MRTrix (26,27). Tract bundles were extracted using an in-house automated tract clustering pipeline (28,29) (Supporting Information Video S1).

Mean ADC (MD) values from each frequency of the multi-frequency scan were obtained in each of the tracts. To mitigate partial volume errors from cerebrospinal fluid that can exhibit negative $\Lambda$ due to flow (9,18), voxels with PGSE MD $> 0.8 \times 10^{-3}$ mm$^2$/s were omitted from the tract-based MD estimates at all frequencies. The primary goal of the multi-frequency scan was to estimate $\theta$ and, because $\theta$ describes a general property of structural organization, we do not expect it to vary across different white matter regions in the healthy brain. Accordingly, to improve robustness against motion, partial volume effects, and Gibbs ringing, a common $\theta$ was assumed over all tracts. $\Lambda$, and $D_{\omega 0}$ were assigned separately to each tract, and all parameters were estimated together using a maximum likelihood estimation of Eq. 1 over the largest 50 tracts in all the subjects, by volume (i.e., 50 tracts per subject). The effective OGSE frequencies used in the fit were estimated as the centroid of the gradient moment power spectrum:

$$\omega = \frac{1}{2\pi} \frac{\int_0^\infty F(\omega)F^*(\omega)\omega d\omega}{\int_0^\infty F(\omega)F^*(\omega)d\omega} \quad [7]$$

where $F(\omega)$ is the Fourier transform of the zeroth moment of the gradient waveform $g(t)$:

$$F(\omega) = \int_{-\infty}^\infty dt\, exp(i\omega t)\int_0^t dt'\gamma g(t') \quad [8]$$

and the acquired signal is related to $F(\omega)$ and the frequency dependent ADC by (9):

$$S = S_0 exp\left(-\frac{1}{\pi}\int_0^\infty F(\omega)D(\omega)F^*(\omega)d\omega\right) \quad [9]$$

***Sequence Optimization***: $T_2$ values from 40 ms to 80 ms were used in the optimization, which covers the range of expected values for both grey and white matter at 7 $T$. The multiple frequency scans implicated $\theta \approx 1/2$ and $\Lambda \approx 10$ μm$^2$/s$^{1/2}$ (see Results below); accordingly, $\theta$ and $\Lambda$ values were estimated to range from 0.4 to 0.6 and 6 to 14 μm$^2$/s$^{1/2}$, respectively. $D_{\omega 0}$ values were estimated to be between 0.6 and 0.8 $\times$ 10$^{-3}$ mm$^2$/s. A maximum gradient amplitude of 68 mT/m and 240 T/m/s slew rate were assumed for simulation, according to limits used experimentally. TE was also calculated from the sequence timings that reflect a 2 mm isotropic in-plane resolution with a single shot EPI readout trajectory, 75% phase-encode partial Fourier, and 2778 Hz/pixel readout bandwidth on our 7 T system. Non-integer values were permitted for the number of periods in the OGSE waveform to avoid discretization of the $\Delta D/\sigma_{\Delta D}$ surface and improve the ability to observe trends in the results. A minimum of 2 OGSE periods was enforced (one on each side of the refocusing RF pulse), because symmetry on either side of the refocusing pulse is required to avoid errors from concomitant gradient fields (30).

***Optimized ADC Difference Mapping Acquisition.*** A dMRI protocol was specified to maximize $\Delta D/\sigma_{\Delta D}$ based on our findings from Eq. 6 (see Results below), and was acquired in the same subjects. This scan consisted of two frequencies acquired with $b = 720$ s/mm$^2$: standard PGSE ($\Delta_{eff} = 32$ ms, 0 Hz), and cosine-modulated trapezoidal OGSE with frequency 38 Hz. The other parameters were 4 direction tetrahedral encoding with 6 averages each, $TE/TR = 82/8200$ ms, $FOV = 200 \times 200$ mm$^2$, 2 mm isotropic in-plane resolution, 48 slices (2 mm), and scan time 6 min. Acquisitions with b = 0 were not acquired.

All image reconstructions used an order 2 Kaiser-Bessel k-space filter to suppress Gibbs ringing (an order 3 filter was used for the multiple frequency scan which had brighter cerebrospinal fluid), PCA denoising (31) before receiver combination, and SENSE-1 coil combination using a direct method that outputs real-valued signal (32).

**Results**

***Multiple Frequency OGSE:*** In the water phantom, ADC values were within 1% of the PGSE value at all OGSE frequencies (Supporting Information Figure S1). MD maps computed from the multiple-frequency scan were of comparable quality over all frequencies (Figure 2) and subjects (Supporting Information Figure S2). Over all subjects, the maximum likelihood estimation of θ was 0.47 ± 0.05, and Λ ranged from 8 ± 4 μm$^2$/s$^{0.53}$ to 13 ± 2 μm$^2$/s$^{0.53}$, depending on the tract (Figure 3). Over all tracts, the mean apparent diffusion dispersion rate was $\Lambda = 11 \pm 1$ μm$^2$/s$^{0.53}$.

***Sequence Optimization*:** Figure 4a shows $\Delta D/\sigma_{\Delta D}$ variation with $b$-value and $\omega$ for $T_2 = 60$ ms, $D_{\omega 0} = 0.7 \times 10^{-3}$ mm²/s, $\theta = 1/2$, and $\Lambda = 10$ μm²/s$^{1/2}$. The minimum required $TE$ for $\Delta D/\sigma_{\Delta D}$ values in Figure 4a are depicted in Figure 4b. Table 1 depicts the optimal acquisition parameters for a range of plausible $T_2$, $D_{\omega 0}$, $\theta$, and $\Lambda$ values. Notably, for all combinations of input parameters, the optimal $\Delta D/\sigma_{\Delta D}$ occurred when only the minimum of 2 OGSE periods were used. Accordingly, Figure 4c depicts the $\Delta D/\sigma_{\Delta D}$ with respect to $\omega$ and $b$ for two periods and the same $T_2$, $D_{\omega 0}$, $\theta$ and $\Lambda$ as in Figure 4a, assuming the gradients are employed at the hardware maximum. Also visible from Table 1 is that the optimal choice of $f$ and $R_{op}$ only weakly depend on the input parameters and are near 40 Hz and 1 for all cases, respectively. Accordingly, $f = 38$ Hz with 2 OGSE periods (corresponding to $b = 720$ s/mm², $TE = 82$ ms, and $R_{op} = 1$) were chosen as parameters for the optimized 6-minute *in vivo* diffusion dispersion scan.

***Optimized ADC Difference Mapping Acquisition*:** Example $\Delta D$ maps computed from the optimized acquisition and corresponding DTI fractional anisotropy (FA) and MD maps are depicted in Figure 5. Example $\Delta D$ maps from multiple slices in all the subjects are shown in Supporting Information Figure S3. Notably, with the assumption of a consistent $\theta$ at all voxels, a $\Lambda$ map can be readily obtained by a simple global scaling of the $\Delta D$ map (Eq. 2); accordingly, a scale bar for $\Lambda$ assuming $\theta = 1/2$ is also shown in Figure 5. Example sagittal images of $\Delta D$ and FA are shown in Figure 6, along with histograms over all subjects of $\Delta D$ (from the optimized scan) and the parallel eigenvalues, perpendicular eigenvalues, and FA (from the DTI scan) in regions-of-interest in the genu, body, and splenium. A trend toward increasing $\Delta D$ from the genu to splenium is observed, in contrast to a U-shaped variation of the diffusion parameters.

**Discussion**

In this work, evidence for a $\omega^{1/2}$ dependence of MD on OGSE frequency was reported for the first time in the human brain in vivo. This trend is similar to recent reports in both healthy (22) and globally ischemic (8,9) rodent brains. In addition, our finding of $\Lambda$ values ~ 10 μm²/s$^{1/2}$ (Figure 3) agrees with ADC results reported at a field strength of 4.7 T (18); for example, $\Lambda$ computed from the corticospinal tract using their reported ADC's is approximately 10 µm$^2$/s$^{1/2}$ when $\theta = 1/2$. Further, optimal methods to acquire maps of the ADC difference between PGSE and OGSE without requiring $b = 0$ images were reported here, which enabled full-brain, clinical resolution maps of the MD difference in a scan time of 6 minutes. The PGSE/OGSE MD difference may be applicable as a new biomarker for pathology and, further, may improve microstructural modelling approaches, as diffusion time dependencies can help resolve model fitting degeneracies (19,33). While the proposed optimized protocol does not provide an estimation of MD, the novel information that is available from OGSE is primarily $\Delta D$ and the acquisition of MD may be better served by a standard DTI acquired with a shorter TE and the same imaging parameters (FOV, resolution, bandwidth). Further, the additional DTI scan would likely already be desired in clinical studies for tractography.

By including $\omega = 0$ in the fitting model, there is an implicit assumption that these experiments were in the long diffusion time regime ($\omega \to 0$), similar to the analysis of rodent data by Novikov (8). In this regime, the observation of $\theta = 1/2$ in white matter is consistent with either highly correlated structural disorder or short-range disorder along one dimension (8). Given that the permeability of myelinated axons is expected to be negligible at these diffusion times (21) and that the intracellular signal is expected to dominate the frequency dependence over extracellular (8), the latter explanation is favored. The assumption of a long diffusion time regime is likely appropriate given that preclinical studies have estimated that frequencies larger than 90 Hz are required to enter the short diffusion time regime for cancer cells that have dimensions > 10 µm (34). Further, in white matter microstructural length scales are smaller (< 10 µm), which would push the required frequency for short diffusion times even higher.

The above hypothesis that the frequency dependence is described by short-range disorder along the axons results in the interpretation that differences in $\Lambda$ (and $\Delta D$ for unchanged $\theta$) describe differences in the amount of disorder along the axons, which may include local variations of thickness or directionality along the axon. This assertion is supported by increases in $\Delta D$ observed acutely after stroke (20), where neurite beading increases disorder along the axons (35). Likewise, the observation in Figure 6 of decreased $\Delta D$ in the genu compared to the splenium is suggestive of less disorder along fibres, which may indicate more consistent axon or other fibre (e.g., astrocyte processes) directionality and/or thickness, or differing volume fractions of axons to support cell processes. Similar large differences of the DTI eigenvalues or FA between the genu and splenium were not observed, which suggests that $\Delta D$ provides complementary microstructural information to DTI.

The primary limitation of this study is that only 4 frequencies is not sufficient for a robust fit of $\theta$. Acquiring data at more frequencies is particularly difficult on human systems due to scan time constraints and because high frequencies drastically reduce the b-value, which in turn reduces the absolute signal differences between different frequencies (because $S_\omega = S_{\omega 0} e^{-b\Delta D}$; see Appendix A). Low non-zero frequencies are also challenging because they require long TE to accommodate a full cosine period on each side of the 180° RF pulse. For example, including a frequency of 15 Hz in the multiple frequency scan would have required a prohibitively long TE of 178 ms. High performance gradient systems with high slew rates and maximum gradient strengths would enable more robust estimation of $\theta$ and $\Lambda$ as they would allow access to higher OGSE frequencies. That said, the findings for optimizing a pulse sequence for measuring the ADC difference between PGSE and OGSE were only weakly dependent on $\theta$.

Our acquisition in a water phantom revealed biases of ADC ~ 0.5% compared to the mean over all frequencies. This may have been caused by eddy current artefacts or slightly nonlinear gradient amplifier gain at very high gradient amplitudes. Nevertheless, these deviations are small compared to the change in ADC of approximately 25% observed in human brain tissue between OGSE at 60 Hz and PGSE. Another potential source of bias for the optimized approach is the omission of $E$ (Eq. 4), which was required to skip $b = 0$ acquisitions. However, for the optimized protocol implemented in this work, this would result in $E \sim 10^{-19}$ mm²/s for $\theta = 1/2$, $\Lambda = 10$ μm²/s$^{1/2}$ and $D_{\omega 0} = 0.7 \times 10^{-3}$ mm²/s; accordingly, $E$ can likely be ignored in practice. Finally, it may also be tempting to further simplify Eq. 4 to use the nominal $b$-value without any consideration of cross-terms; however, this would lead to an error in $\Delta D$ of approximately 7% and is accordingly not recommended.

The $\Delta D/\sigma_{\Delta D}$ computations suggested that an OGSE frequency ~ 40 Hz is optimal for a broad range of physiologically feasible values of $T_2$, $D_0$, and the diffusion dispersion power law scaling ($\theta$) and rate ($\Lambda$). However, the optimal values strongly depend on gradient hardware limits or direction schemes. For example, the optimal frequency for tetrahedral encoding with 300 mT/m gradients is 100 Hz (all other parameters, including slew rate, kept the same as those used for Table 1). Notably, in this case the optimal parameters are still obtained with only 2 OGSE periods, similar to the results here. This general finding suggests that increasing the $b$-value by increasing the number of OGSE periods is not worth the SNR losses incurred by the greatly increased TE. This conclusion does not consider the narrowing of OGSE spectra that occurs with an increased number of periods; however, given the generally low $\Delta D/\sigma_{\Delta D}$ achievable on human systems, remedying this spectral blurring may not be worth the $\Delta D/\sigma_{\Delta D}$ cost.

The noise propagation analysis did not consider Rician noise because real-valued images with Gaussian noise were used in this work. However, the results likely approximately apply to absolute value images with Rician noise because achieving $\Delta D$ maps with reasonable SNR requires the raw signal level for both OGSE and PGSE to be much higher than the noise floor (e.g., the optimal $\Delta D/\sigma_{\Delta D}$ < 5% of the PGSE signal; Figure 4).

The optimized acquisition did not acquire any imaging volumes with $b = 0$, and the diffusion weighted images at the various frequencies were compared directly using Eq. 4 with $E$ ignored. Notably, $b = 0$ images have extremely bright CSF, particularly at the long TE required for OGSE, which results in severe Gibb's ringing. When an MD is computed, the different Gibbs ringing profiles for $b = 0$ and diffusion weighted acquisitions causes amplified ringing in MD maps (36). Because the computation of $\Delta D$ in Eq. 4 compares only diffusion weighted signals with the same diffusion weighting, the CSF signal is fairly consistent, and this type of Gibbs ringing amplification is partially mitigated (Figure 7). That said, the CSF signal for PGSE is lower than for OGSE because of signal losses from incoherent flow (OGSE is inherently flow-compensated), which results in negative $\Delta D$ values in the fluid (9,18). Negative $\Delta D$ is not physiologically plausible in brain tissue, where diffusion is restricted/hindered and flow is absent (33); accordingly, voxels with negative $\Delta D$ were masked in displayed images.

OGSE acquisitions on human systems utilize relatively low $b$-values, which may make estimations of diffusion dispersion sensitive to perfusion. For the acquisitions and fitted parameters to be insensitive to perfusion, the perfusion signal must be much less than the tissue signal at all frequencies used. In mice, no dependence on perfusion was observed for frequencies up to 200 Hz for $b > 300$ s/mm² (22); accordingly, this assumption was likely satisfied here, where $b$-values of at least 450 s/mm² and a maximum frequency of only 60 Hz were utilized. On the other hand, at high b-values, higher order terms in the cumulant expansion of $D(\omega)$ (16) and rotational variance (37,38) may need to be considered for accurate estimation of $\theta$ and $\Lambda$. However, at the b-values used here ($\leq 700$ s/mm²), it is not expected that higher order terms or rotational variance affected our results (39).

While $\theta = 1/2$ was implicated here in healthy human white matter, this may not be the case in other tissue types or in neurological disorders. Accordingly, care should be taken in the interpretation of changes in $\Delta D$ in pathology, which could result from a combination of $\theta$ and $\Lambda$ changes.

**Conclusion**

In conclusion, we have provided evidence for a $\omega^{1/2}$ dependence of ADC in the *in vivo* human brain using oscillating gradient spin-echo diffusion MRI and developed an optimized acquisition protocol that enabled full brain mapping of ADC differences between PGSE and OGSE in a clinically relevant 6 minutes. The ability to rapidly probe diffusion dispersion in vivo opens the door for the exploration of new biomarkers and more sophisticated microstructural models.

**Acknowledgements**

The authors thank the Canada first research excellence fund to BrainsCAN, Western Strategic Support Program, and the Natural Sciences and Engineering Research Council of Canada.

**Appendix A**

Neglecting the signal bias from omitting the $b = 0$ scan, the difference in mean ADC between OGSE and PGSE is, from Eq. 4:

$$\Delta D = \frac{1}{N_\omega} \sum_{i=1}^{N_\omega} \frac{-\ln(S_{\omega,i})}{b_{\omega,i}} + \frac{1}{N_{\omega 0}} \sum_{i=1}^{N_{\omega 0}} \frac{\ln(S_{\omega 0,i})}{b_{\omega 0,i}} \qquad [10]$$

where $S_{\omega,i}$ and $S_{\omega 0,i}$ are the OGSE and PGSE signals along the different encoding directions $i$, respectively, $b$ is the diffusion weighting, $\omega^\theta$ is the frequency, $N_\omega$ is the number of OGSE acquisitions, and $N_{\omega 0}$ is the number of PGSE acquisitions. To simplify variance propagation with a modest loss of generality, we will assume equal signal levels and $b$-values along all directions (i.e., isotropic diffusion and negligible gradient cross terms) and the same noise variance, $\sigma$, for both PGSE and OGSE, which yields:

$$\Delta D = \frac{-\ln(S_\omega)}{b} + \frac{\ln(S_{\omega 0})}{b}, \qquad [11]$$

$$\sigma^2_{\Delta D} = \frac{1}{N_\omega}\left(\frac{\partial \Delta D}{\partial S_\omega}\right)^2 \sigma^2 + \frac{1}{N_{\omega 0}}\left(\frac{\partial \Delta D}{\partial S_{\omega 0}}\right)^2 \sigma^2$$

$$= \frac{\sigma^2}{b^2}\left(\frac{1}{N_{\omega 0} S^2_{\omega 0}} + \frac{1}{N_\omega S^2_\omega}\right) \qquad [12]$$

where $\sigma^2_{\Delta D}$ is the variance in $\Delta D$. Substituting $S_\omega = S_{\omega 0} e^{-b\Delta D}$ (obtained from rearranging Eq. 11) into Eq. 12 and setting the total number of acquisitions $N = N_{\omega 0} + N_\omega$ yields:

$$\sigma^2_{\Delta D} = \frac{\sigma^2}{S^2_{\omega 0} b^2}\left(\frac{1}{N_{\omega 0}} + \frac{e^{2b\Delta D}}{N - N_{\omega 0}}\right) \qquad [13]$$

The relationship between $S_{\omega 0}$ and the sequence parameters $b$ and $TE$, can be represented by $S_{\omega 0} = Ce^{-bD_{\omega 0}}e^{-TE(b)/T_2}$, where C is a constant related to receiver sensitivities and proton density and $D_{\omega 0}$ is the ADC measured using PGSE. The function $TE(b)$ depends on $b$ and also implicitly depends on the gradient hardware limits and EPI timings. For OGSE, the b-value is increased by adding more periods on each side of the 180° RF pulse, which in turn lengthens the TE. Substituting this expression for $S_{\omega 0}$ into Eq. 13 and taking the square root yields:

$$\sigma_{\Delta D} = \frac{\sigma}{Cb}\left(\frac{1}{N_{\omega 0}} + \frac{e^{2b\Delta D}}{N-N_{\omega 0}}\right)^{\frac{1}{2}} e^{bD_{\omega 0}} e^{TE(b)/T_2} \qquad [14]$$

We define the SNR of the proton density image with $TE = 0$ as $SNR_0 = C/\sigma$, which results in the following expression for the signal-to-noise ratio of the ADC difference:

$$\frac{\Delta D}{\sigma_{\Delta D}} = SNR_0 \cdot b\Delta D\left(\frac{1}{N_{\omega 0}} + \frac{e^{2b\Delta D}}{N-N_{\omega 0}}\right)^{-\frac{1}{2}} e^{-bD_{\omega 0}} e^{-TE(b)/T_2} \qquad [15]$$

Determination of the optimal OGSE frequency based on maximizing $\Delta D/\sigma_{\Delta D}$ requires knowledge of the dependence of $\Delta D$ on $\omega$. Assuming a power law relationship, $\Delta D = \Lambda\omega^\theta$, yields:

$$\frac{\Delta D}{\sigma_{\Delta D}} = SNR_0 \cdot b\Lambda\omega^\theta\left(\frac{1}{N_{\omega 0}} + \frac{e^{2b\Lambda\omega^\theta}}{N-N_{\omega 0}}\right)^{-\frac{1}{2}} e^{-bD_{\omega 0}} e^{-TE(b)/T_2} \qquad [16]$$

The optimal choices of $b$, $\omega$, and $N_{\omega 0}$ can be straight-forwardly determined numerically by performing an exhaustive search to find the combination that maximizes $\Delta D/\sigma_{\Delta D}$. However, more insight into the optimal $N_{\omega 0}$ can be obtained by taking the partial derivative with respect to $N_{\omega 0}$ and setting the result to zero:

$$-\frac{1}{2}\left(\frac{1}{N_{\omega 0}} + \frac{e^{2b\Lambda\omega^\theta}}{N-N_{\omega 0}}\right)^{-\frac{3}{2}} \left(-\frac{1}{N_{\omega 0}^2} + \frac{e^{2b\Lambda\omega^\theta}}{(N-N_{\omega 0})^2}\right) = 0 \qquad [17]$$

Recognizing that the ratio of OGSE acquisitions to PGSE acquisitions is $R_{op} = (N - N_{\omega 0})/N_{\omega 0}$, Eq. 17 simplifies to

$$R_{op} = e^{b\Lambda\omega^\theta} \qquad [18]$$

This motivates rearranging Eq. 16 to include $R_{op}$:

$$\frac{\Delta D}{\sigma_{\Delta D}} = SNR_0\sqrt{N_{\omega 0}} \cdot b\Lambda\omega^\theta\left(1 + \frac{e^{2b\Lambda\omega^\theta}}{R_{op}}\right)^{-\frac{1}{2}} e^{-bD_{\omega 0}} e^{-TE(b)/T_2} \qquad [19]$$

It is worth noting that, since the rate of diffusion dispersion ($\Lambda$) is directly proportional to $\Delta D$, the signal-to-noise ratio of an estimation of $\Lambda$ is equivalent to the expression given in Eq. 19 for an *a priori* known $\theta$ (i.e., $\Delta D/\sigma_{\Delta D} = \Lambda/\sigma_\Lambda$).

Clustering of White Matter Tractography. In: Joint Annual Meeting ISMRM-ESMRMB 2018.

29. Kai J, Khan AR. NeuroBundle Extraction and Evaluation Resource.; 2019. doi: 10.5281/zenodo.3350903.

30. Baron CA, Lebel RM, Wilman AH, Beaulieu C. The effect of concomitant gradient fields on diffusion tensor imaging. Magn. Reson. Med. 2012;68:1190–1201.

31. Veraart J, Novikov DS, Christiaens D, Ades-Aron B, Sijbers J, Fieremans E. Denoising of diffusion MRI using random matrix theory. Neuroimage 2016;142:394–406.

32. McKenzie CA, Yeh EN, Ohliger MA, Price MD, Sodickson DK. Self-calibrating parallel imaging with automatic coil sensitivity extraction. Magn. Reson. Med. 2002;47:529–538.

33. Novikov DS, Fieremans E, Jespersen SN, Kiselev VG. Quantifying brain microstructure with diffusion MRI: Theory and parameter estimation. NMR Biomed. 2018:e3998.

34. Reynaud O, Winters KV, Hoang DM, Wadghiri YZ, Novikov DS, Kim SG. Surface‑to‑volume ratio mapping of tumor microstructure using oscillating gradient diffusion weighted imaging. Magn. Reson. Med. 2016;76:237–247.

35. Budde MD, Frank JA. Neurite beading is sufficient to decrease the apparent diffusion coefficient after ischemic stroke. Proc. Natl. Acad. Sci. U. S. A. 2010;107:14472–14477.

36. Veraart J, Fieremans E, Jelescu IO, Knoll F, Novikov DS. Gibbs ringing in diffusion MRI. Magn. Reson. Med. 2016;76:301–314.

37. Nørhøj Jespersen S, Lynge Olesen J, Ianuş A, Shemesh N. Effects of nongaussian diffusion on "isotropic diffusion" measurements: an ex-vivo microimaging and simulation study. J. Magn. Reson. 2019 doi: 10.1016/j.jmr.2019.01.007.

38. Henriques RN, Jespersen SN, Shemesh N. Microscopic anisotropy misestimation in spherical-mean single diffusion encoding MRI. Magn. Reson. Med. 2019;81:3245–3261.

39. Nilsson M, Szczepankiewicz F, Brabec J, et al. Tensor‑valued diffusion MRI in under 3 minutes: an initial survey of microscopic anisotropy and tissue heterogeneity in intracranial tumors. Magn. Reson. Med. 2019;2:581.
20

**Tables**

*Table 1*: $\Delta D/\sigma_{\Delta D}$ optimization.

| Tissue Properties | | | | Optimal Diffusion Encoding Parameters | | | |
|---|---|---|---|---|---|---|---|
| $T_2$ [ms] | $\Lambda$ [$\mu m^2/s^{1-\theta}$] | $D_{\omega 0}$ [$\mu m^2/ms$] | $\theta$ | $\omega/2\pi$ [Hz] | $b$-value [s/mm$^2$] | Min $TE$ [ms] | $R_{op}$ |
| 60 | 10 | 0.7 | 0.5 | 39 | 700 | 81 | 1.12 |
| 40 | 10 | 0.7 | 0.5 | 43 | 520 | 76 | 1.09 |
| 80 | 10 | 0.7 | 0.5 | 38 | 755 | 82 | 1.12 |
| 60 | 6 | 0.7 | 0.5 | 38 | 755 | 82 | 1.07 |
| 60 | 14 | 0.7 | 0.5 | 40 | 650 | 80 | 1.16 |
| 60 | 10 | 0.6 | 0.5 | 38 | 755 | 82 | 1.12 |
| 60 | 10 | 0.8 | 0.5 | 41 | 600 | 78 | 1.10 |
| 60 | 10 | 0.6 | 0.4 | 38 | 755 | 82 | 1.08 |
| 60 | 10 | 0.8 | 0.6 | 41 | 600 | 78 | 1.15 |

Optimal diffusion encoding parameters ($\omega$, $b$-value, $TE$, and $R_{op}$) vary with $T_2$, $D_{\omega 0}$, $\theta$ and $\Lambda$. The optimal parameters for the nominal expected parameters are shown in the top row, while the other rows show the result of varying the highlighted cells with respect to the top row values. Notably, $\omega$ and $R_{op}$ are almost independent on the input parameters.

**Figures**

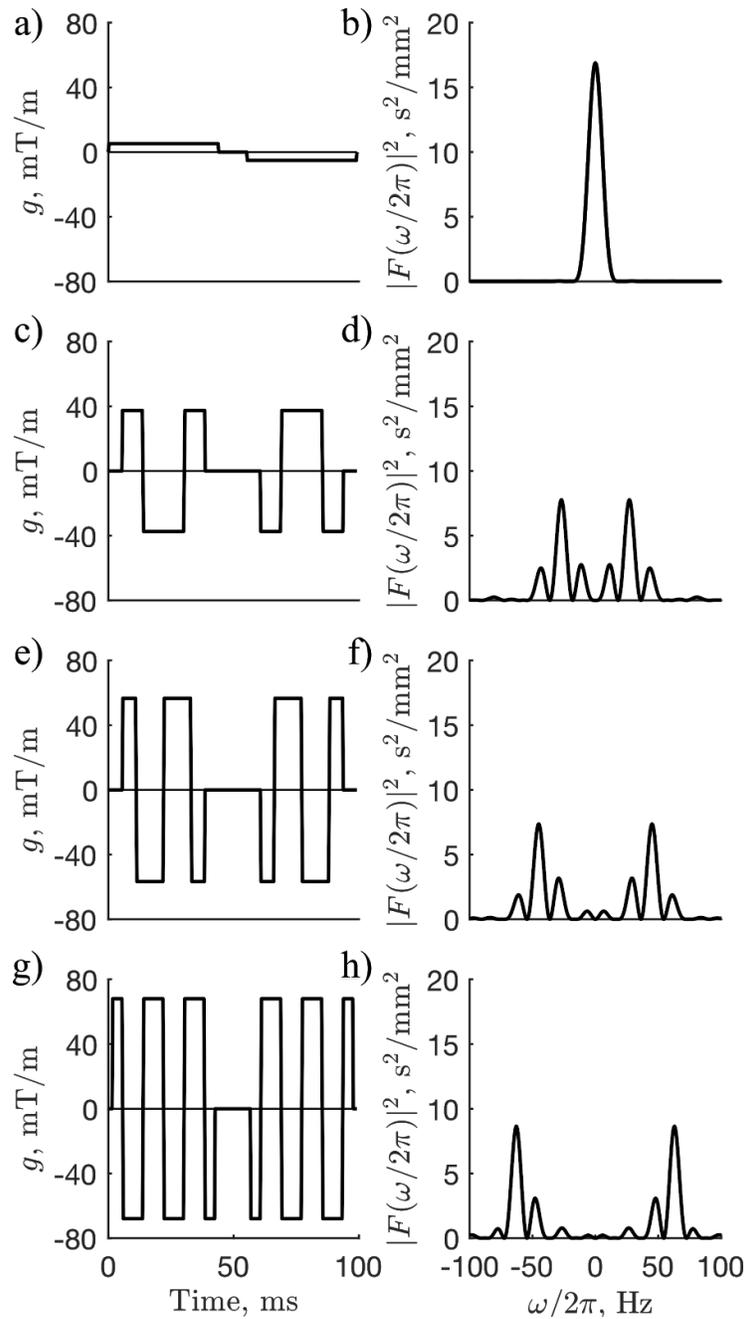

**Figure 1:** Gradient waveforms and frequency spectra (Eq. 8) for the multi-frequency scan, which used nominal frequencies of 0 Hz (a, b), 30 Hz (c, d), 45 Hz (e, f), and 60 Hz (g, h). Implicit gradient reversal due to the 180° RF pulse has been applied, and the shown gradient amplitudes were applied simultaneously on all 3 gradient channels.

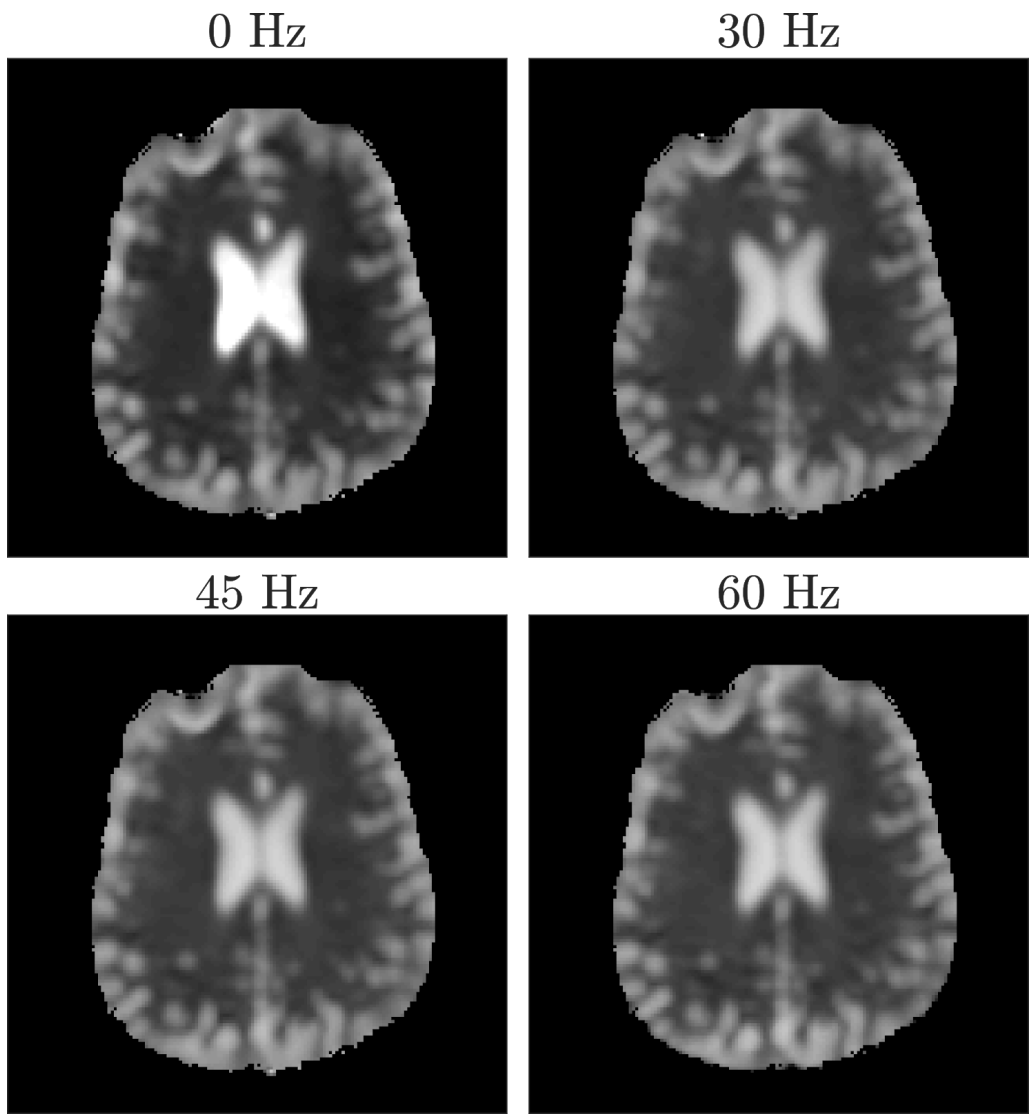

**Figure 2**: Example MD maps in one subject where comparable image quality is observed across OGSE frequencies.

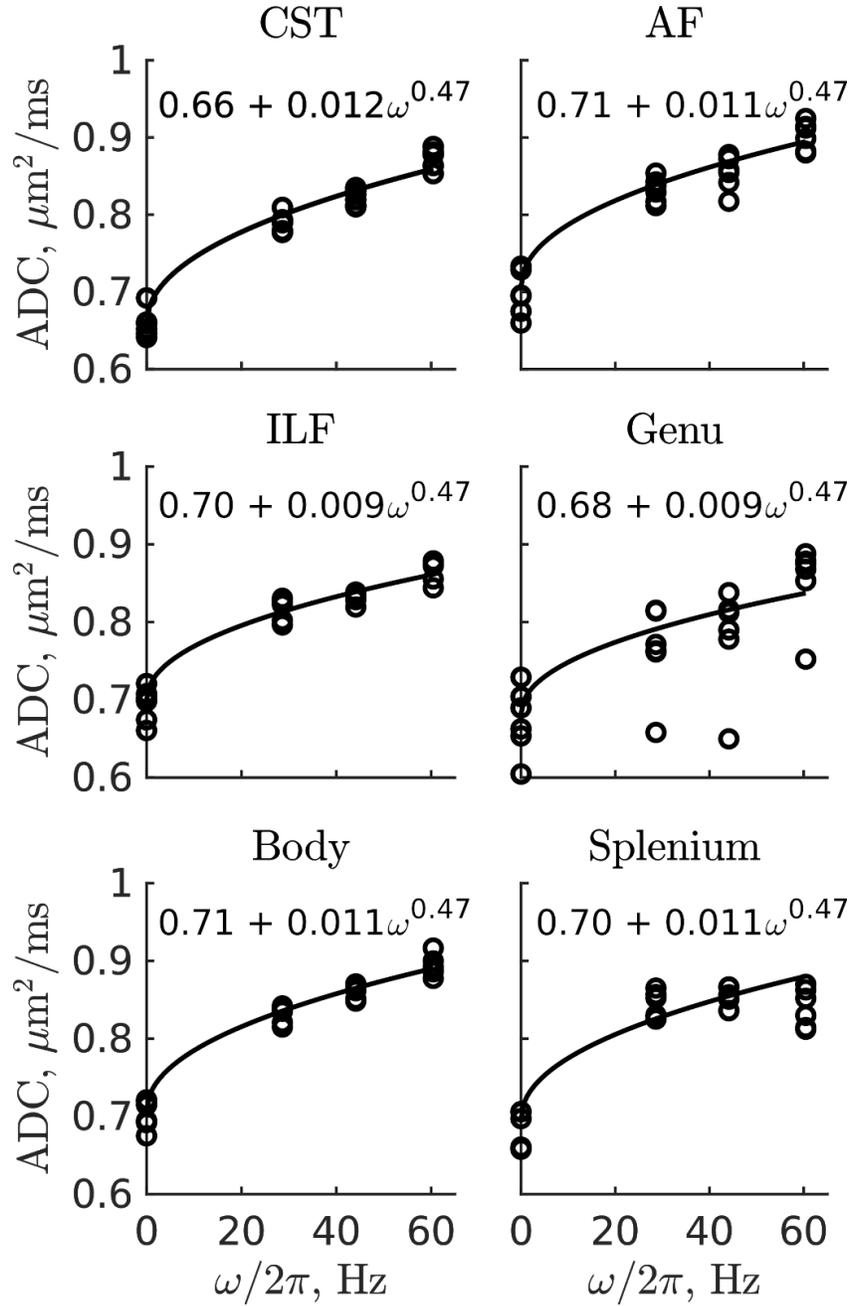

**Figure 3**: Mean ADC values fitted to a power law in a subset of the 50 tracts analyzed (each data point is the mean value within the tract in each subject), where $\theta$ was assumed to be consistent across all tracts. $\theta = 0.47 \pm 0.05$ and $\Lambda$ ranged from $8 \pm 4$ $\mu m^2/s^{0.53}$ to $13 \pm 2$ $\mu m^2/s^{0.53}$, depending on the tract. Over all tracts, the mean apparent diffusion dispersion was $\Lambda = 11 \pm 1$ $\mu m^2/s^{0.53}$. CST - corticospinal tract; AF - arcuate fasciculus; ILF - inferior longitudinal fasciculus; Genu, Body, Splenium - genu, body, and splenium of the corpus callosum.

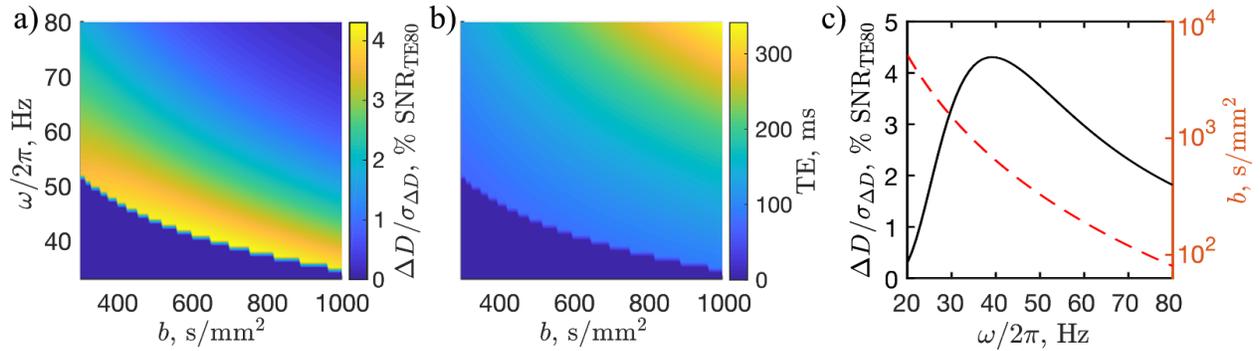

**Figure 4**: $\Delta D/\sigma_{\Delta D}$ optimization: (a) $\Delta D/\sigma_{\Delta D}$ variation with $\omega$ and $b$-value for $T_2$ = 60 ms, $D_{\omega 0}$ = 0.7 × 10$^{-3}$ mm²/s, $\theta = 1/2$, and $\Lambda = 10$ µm²/s$^{1/2}$, (b) required minimum $TE$ for $\Delta D/\sigma_{\Delta D}$ values in (a), and (c) $\Delta D/\sigma_{\Delta D}$ variation with $\omega$ and $b$-value with the total number of OGSE periods fixed at 2 (one on each side of the refocusing RF pulse), which also monotonically links the b-value to the OGSE frequency. In (c), $\Delta D/\sigma_{\Delta D}$ was maximized at a frequency of 39 Hz (i.e., $\omega$ = 245 rad/s) and $b$ = 700 s/mm² with minimum required $TE$ = 81 ms. The lower-left region where $\Delta D/\sigma_{\Delta D} = 0$ in (a) and (b) corresponds to experimentally impossible diffusion encoding states requiring fewer than 2 OGSE periods in total. $\Delta D/\sigma_{\Delta D}$ is displayed as a percentage of the SNR of a $b = 0$ scan with TE = 80 ms and $N_{\omega 0}$ averages.

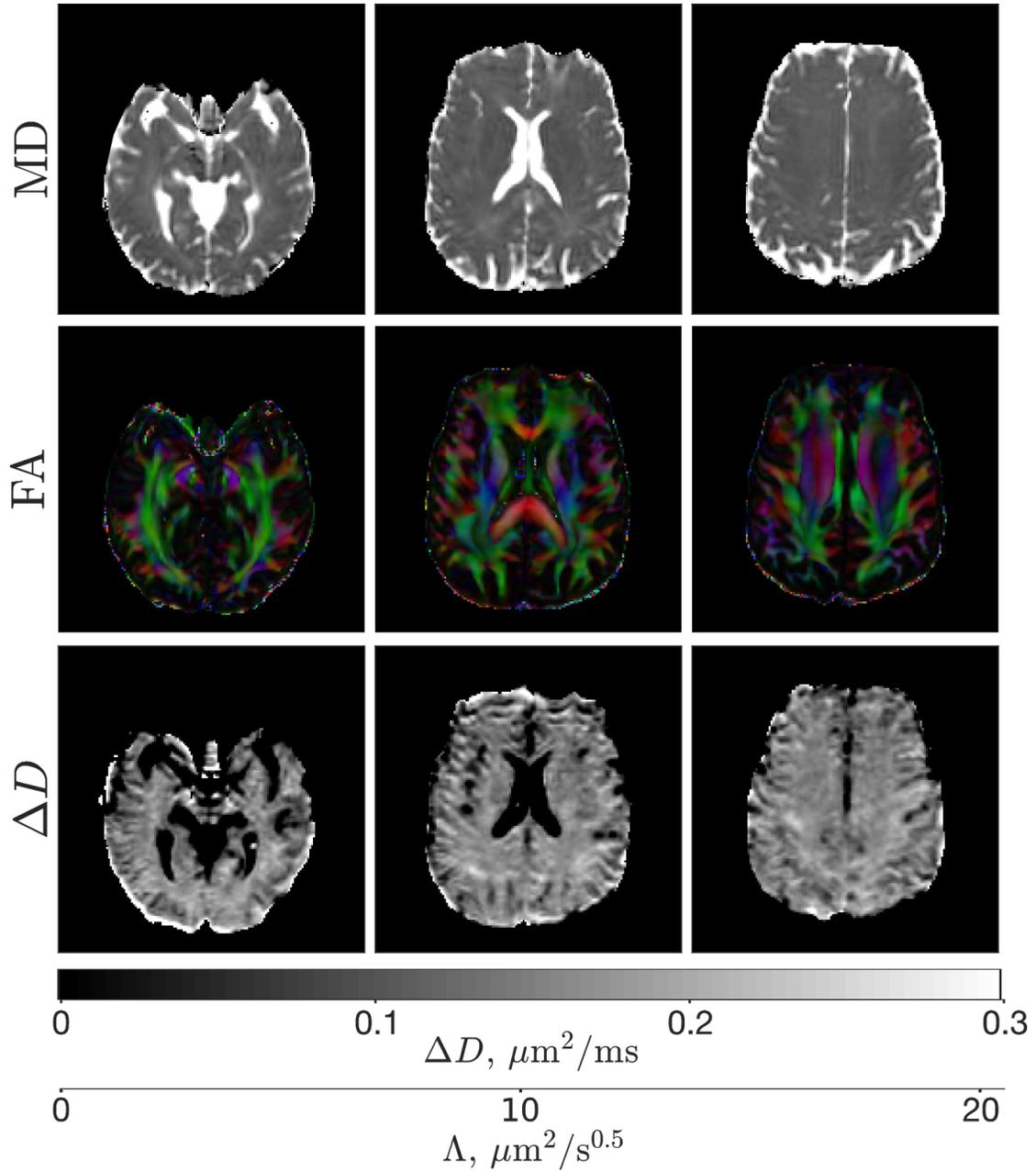

**Figure 5**: Example MD, color-coded FA, and optimized $\Delta D$ maps from multiple slices in one subject. MD and FA maps were computed from the standard DTI scan, and $\Delta D$ maps were calculated from the optimized acquisition. A scale bar for $\Lambda$ is also shown for the case when $\theta = 1/2$. Voxels with negative $\Lambda$ are set to zero in the $\Delta D$ maps, which generally occurs in the cerebrospinal fluid.

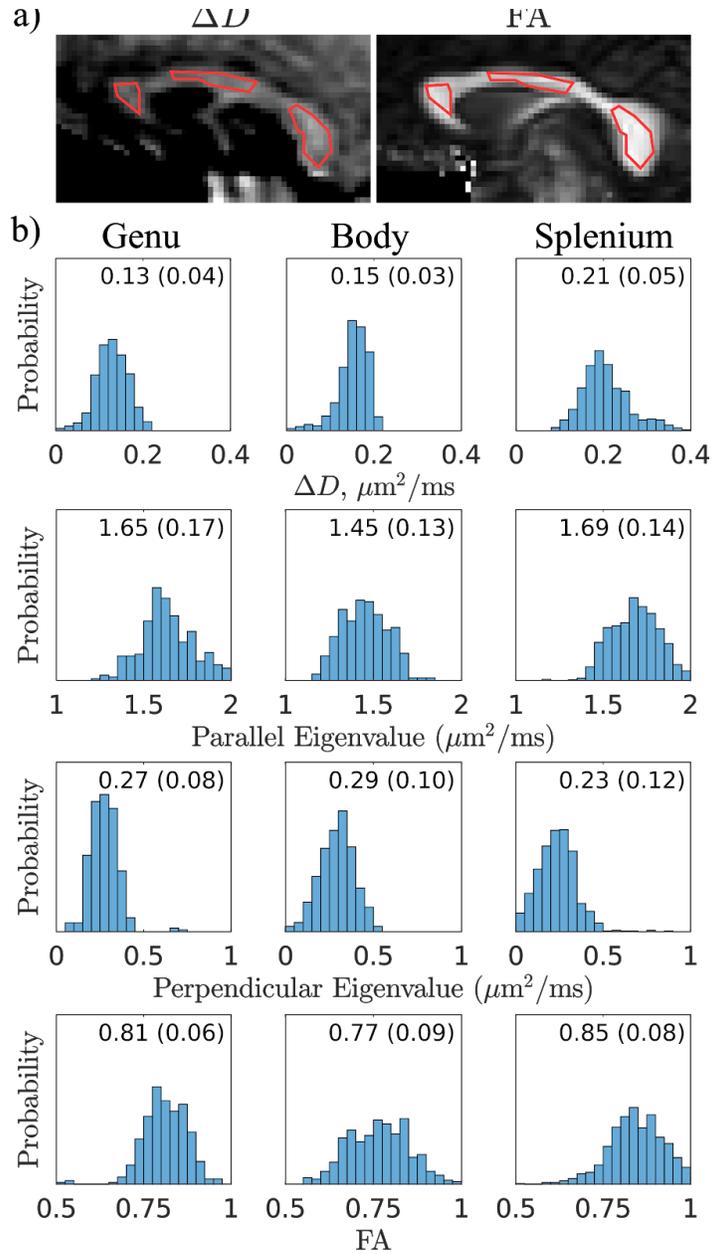

**Figure 6:** (a) Optimized $\Delta D$ map and FA map (latter from DTI scan) in the corpus callosum of one subject. A trend of increasing $\Delta D$ is observed from the genu to the splenium. (b) Histograms over all voxels and subjects of $\Delta D$ from the optimized diffusion dispersion scan and parallel eigenvalue, perpendicular eigenvalue, and FA from the DTI scan are shown in the genu, body of the corpus callosum, and splenium. The values shown near the top of each plot indicate the mean (standard deviation) of the voxel-wise values in the histogram. The regions-of-interest in (a) indicate example regions that were used for the histograms.

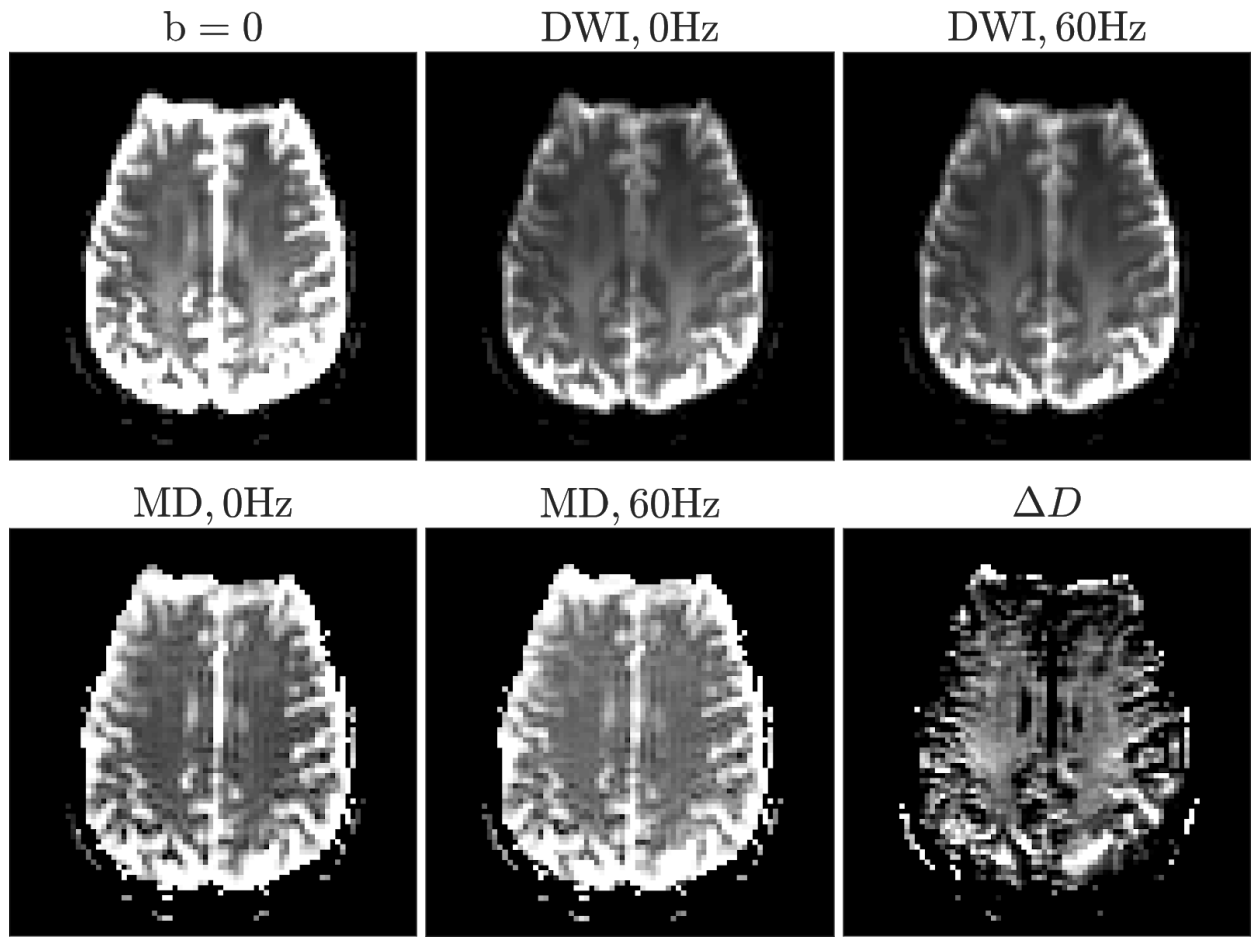

**Figure 7**: (Top row) Raw data, from left to right: non-diffusion weighted image, PGSE diffusion weighted image (DWI), and 60 Hz OGSE DWI. The CSF signal is greatly reduced in the DWIs compared to the non-diffusion weighted image. (Bottom row) MD and $\Delta D$ maps, from left to right: PGSE MD, 60 Hz OGSE MD, and $\Delta D$ maps. Notably, less ringing is observed for $\Delta D$ compared to MD. Images are shown from an individual slice of the multiple-frequency scan in one subject, but similar trends were generally observed in all subjects.

**Supporting Information Video S1**: The largest 50 tracts, by volume, generated by tract clustering in one subject.

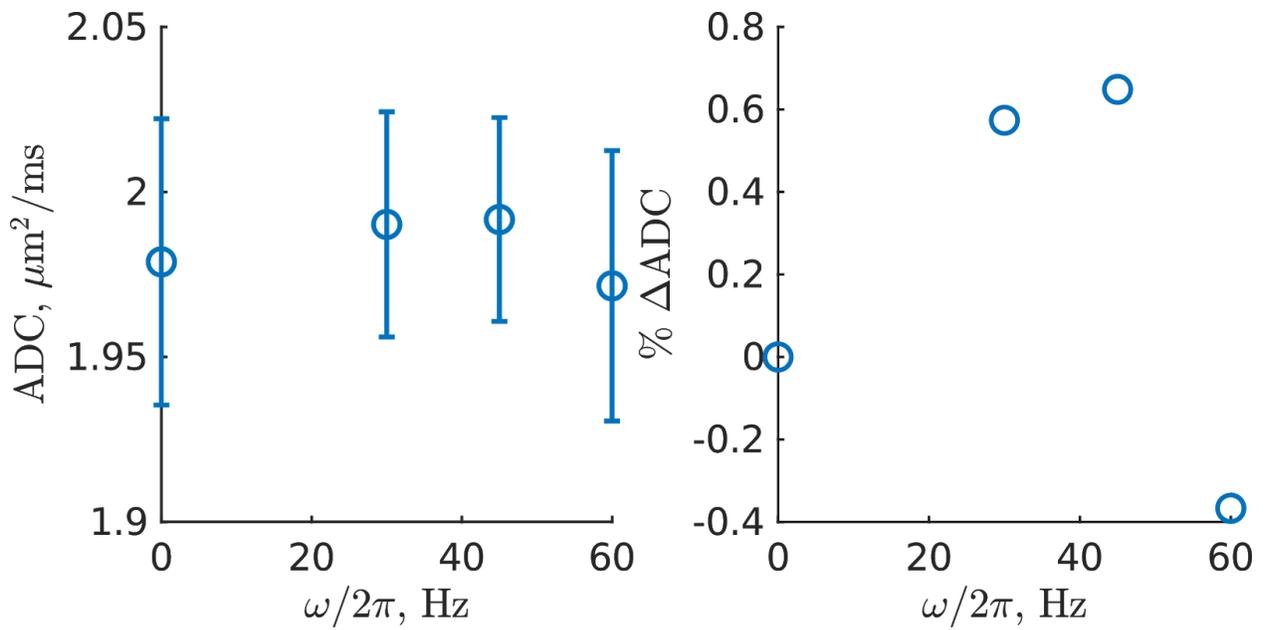

**Supporting Information Figure S1**: Frequency dependence of ADC in a water phantom without diffusion restriction: (a) Mean ADC at 0 Hz (PGSE), 30 Hz, 45 Hz, and 60 Hz (OGSE), (b) Percent change in ADC across the chosen encoding frequencies with respect to the reference PGSE ADC. Mean ADC for each frequency was calculated within a large multi‑slice ROI covering a 24 mm thick slab centered at isocenter (12 slices).

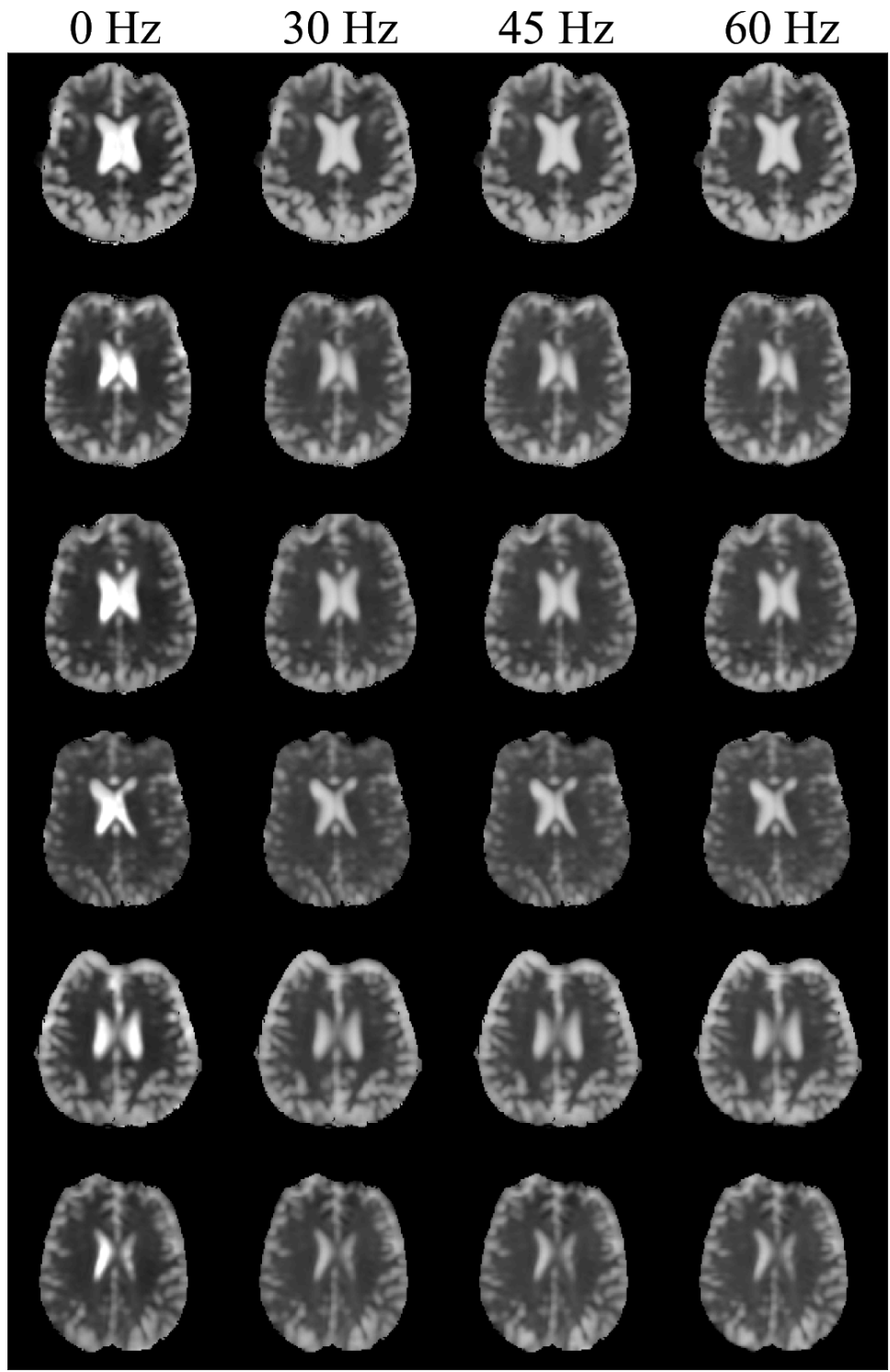

**Supporting Information Figure S2:** Example MD maps from the multiple‑frequency scan in all 6 subjects, from left to right: PGSE ($\Delta_{eff}$ = 41 ms, 0 Hz), 30 Hz OGSE, 45 Hz OSGE, 60 Hz OGSE.

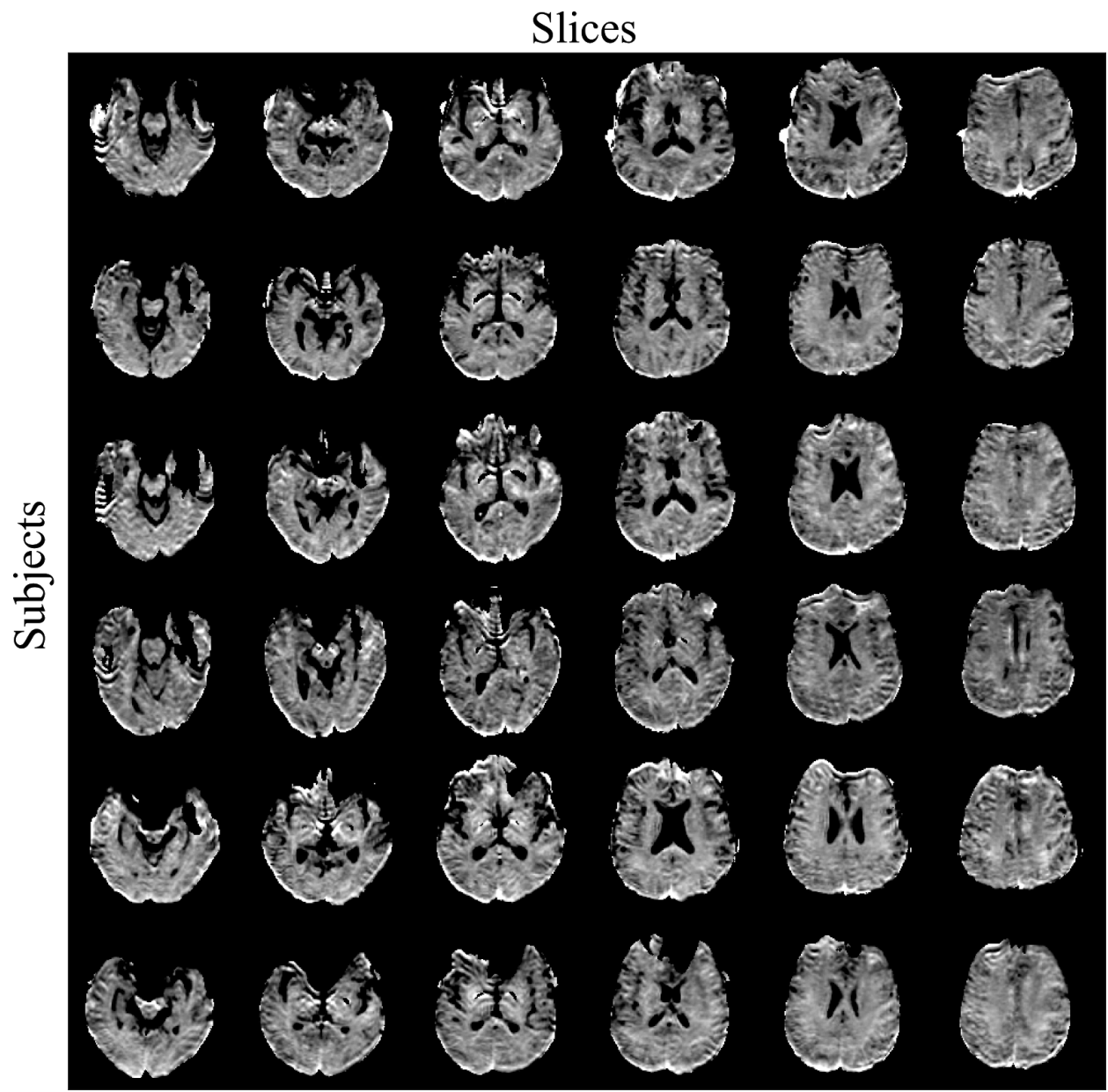

**Supporting Information Figure S3:** Example $\Delta D$ maps from the optimized 6-minute scan in multiple slices of all 6 subjects.